# A Pilot Study of Smart Agricultural Irrigation using Unmanned Aerial Vehicles and IoT-Based Cloud System

**Mohamed Esmail Karar** [1,2], **Faris Alotaibi** [1], **Abdullah AL Rasheed** [1], **Omar Reyad** [1,3]

[1] College of Computing and Information Technology, Shaqra University, Saudi Arabia

[2] Faculty of Electronic Engineering, Menoufia University, Egypt

[3] Faculty of Science, Sohag University, Egypt

*mkarar@su.edu.sa* ; *S437460268@su.edu.sa* ; *S437460255@su.edu.sa* ; *oreyad@su.edu.sa*

**Abstract**—This article introduces a new mobile-based application of modern information and communication technology in agriculture based on Internet of Things (IoT), embedded systems and an unmanned aerial vehicle (UAV). The proposed agricultural monitoring system was designed and implemented using Arduino microcontroller boards, Wi-Fi modules, water pumps and electronic environmental sensors, namely temperature, humidity and soil moisture. The role of UAV in this study is to collect these environmental data from different regions of the farm. Then, the quantity of water irrigation is automatically computed for each region in the cloud. Moreover, the developed system can monitor the farm conditions including the water requirements remotely on Android mobile application to guide the farmers. The results of this study demonstrated that our proposed IoT-based embedded system can be effective to avoid unnecessary and wasted water irrigation within the framework of smart agriculture.

**Keywords**—Internet of things, precision agriculture, embedded systems, cloud computing, smart irrigation, Agricultural drone

1. Introduction

Water is one of essential natural resources for agriculture. Most growing population in the word is relying on agriculture and consumes about 56.0 % of the fresh water globally [1]. However, water availability for irrigation is currently limited because of climate change and other environmental factors such as decreasing groundwater resources, while increasing industrial and domestics demands [2]. Additionally, a large amount of irrigation water is always wasted and cannot be managed efficiently because of traditional irrigation methods to achieve expected crop productivity. Therefore, design and development of smart water irrigation systems become an attractive topic for many researchers because of its potential outcome to achieve sustainable agriculture [3, 4].

Smart agriculture or precision agriculture provides an application of modern and accurate information and communications technology to automate and optimize difficult agricultural tasks and/or processes [5, 6]. Internet of Things (IoT) presents the core networking technology of smart agriculture and many recent wireless applications ranging from environment monitoring to healthcare and medical applications [7-11]. In smart agriculture, IoT is used to collect data from sensors which measure different parameters such as soil moisture and humidity and monitor them remotely via developed mobile applications to take decisions or to actuate devices, e.g., water pumps [6]. Consequently, the workload of the farmers will be reduced with efficient use of available resources and low-cost appliances. Artificial intelligence (AI) techniques can be also applied to analyze agricultural and environmental data to support the specialists and famers; for example, machine learning including neural networks [12], fuzzy logic controllers [13], and recently deep learning models [14]. However, continuous screening and monitoring of the crops, automated water irrigation, and plant disease detection are still main challenges of smart agriculture, as depicted in Fig. 1 [6, 15]. In this study, we will focus only on proposing two solutions for the challenges of continuous monitoring of the field and automatic irrigation using IoT and unmanned aerial vehicles (UAVs).

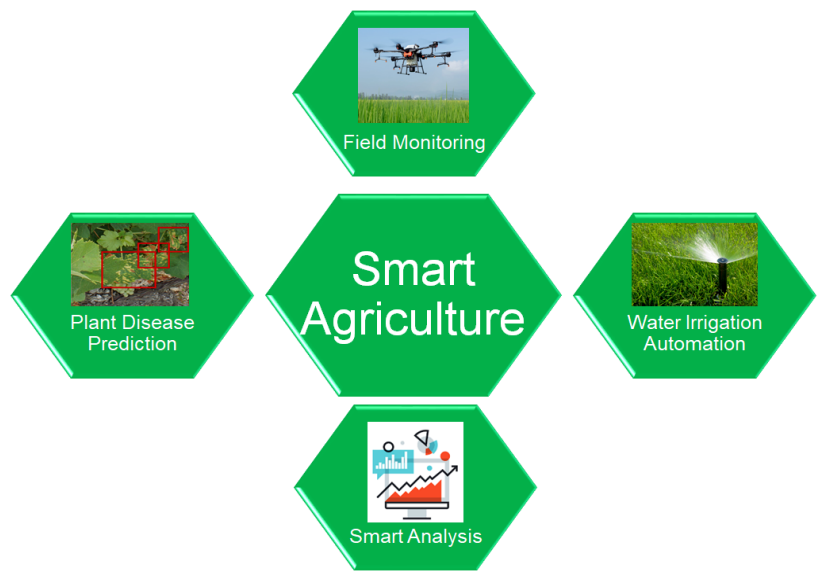

**Fig.1 Main challenges of smart agriculture**

The UAVs or unmanned aircraft system (UASs) are also known as drones. Essentially, a drone is a flying robot that can be remotely controlled. In past years, UAVs were often associated with the military applications, where they were used initially for anti-aircraft target practice, intelligence gathering and used as weapons. Recently, drones are used in a wide range of civilian roles and applications such as rescue, surveillance, traffic monitoring, weather monitoring, and delivery services [16, 17].

Many studies highlight the role of digitally controlled farm plants, i.e., UAVs for monitoring and forecasting in agriculture to keep health conditions of the crops [18, 19]. Drones are also capable of accomplishing intelligent irrigation monitoring using thermal or infrared imaging cameras in Cloud of Things (CoT) network [20]. In addition, manual spray of the pesticides in the filed causes dangerous diseases for the farming workers as reported globally by the Food and Agriculture Organization (FAO) and the World Health Organization (WHO) [21]. Therefore, the UAVs represent a good solution to automatically spray the pesticides, and to minimize potential environmental and health problems of the farmers [18, 19]. Fig. 2 shows the basic components of UAVs, including the frame, propellers, brush-less motors, Electronic Speed Control (ESC) modules, flight control board, RF transmitter and receiver modules, and rechargeable battery [22].

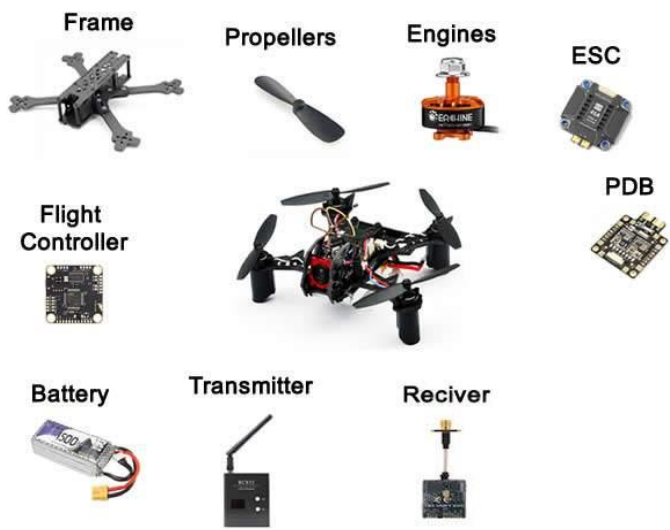

**Fig.2 Basic components of a drone**

**1.1 Related Work**

For smart farming purposes, the application of UAV technology with IoT-based environmental sensor networks have played an important role to enhance the productivity of agricultural crops in the last years [23-25]. This role of UAVs along with IoT-based embedded systems for water irrigation management in smart agriculture has been highlighted in previous studies as follows. Chebrolu *et al*. [26] proposed a registration method of UAV images to reconstruct a three-dimensional (3D) model of the crop. That allows monitoring of the crop growth parameters based on a plant-level. Similarly, the height of maize and Sorghum plants in the field has been determined based on the UAV images and 3D reconstructed plant model [27]. The results of this study showed that the Root Mean Square Error (RMSE) was 0.33 m for the average single Sorghum height. Also, leaf area index (LAI) in Soybean plants has been extracted using UAV and 3D plant model [28]. A comparative study of using different cameras mounted on the UAV for precision agriculture has been conducted in [29]. The authors verified that UAV-based multi-spectral images are instructive and can constitute a great potential for smart agriculture. Based on visible reflectance for evaluating crop biomass, vegetation indices can be extracted using RGB camera with a drone within an acquisition platform [30, 31]. A case study in Indonesia [32] has been conducted to demonstrate the potential of deploying UAV-based remote sensing to assist precision agriculture mapping. That includes collecting periodic information of the field, such as vegetation monitoring, plant healthy, and stock valuation.

Many researchers investigated the feasibility of employing IoT technology systems to control water irrigation and crops monitoring. Automated water irrigation was introduced and implemented using a mobile application [33]. The designed smartphone application can acquire and process images of the soil around the root area of the plants to determine sensor-less the current quantity of water. Based on ARM9 processor, a smart drip irrigation system was developed, displaying environmental conditions such as high temperature, low moisture, $CO_2$ concentration, and the activities of the developed system [34]. Zaier et al. [35] proposed a wireless-controlled irrigation system to leverage the use of groundwater in the large-scale farms of Oman. Zhang *et al*. [36] exploit the IoT and big data to develop a real-time monitoring system of the growth of crops to support the water irrigation decisions. They proposed a central unit to develop a crop growth model based on extracted plant data. The Open Platform Communications (OPC) have been investigated for smart farming. The feasibility of OPC unified architecture was further studies to be deployed for mobile applications of agricultural machinery systems in [37]. To monitor specific crop conditions, Zhang et al. [38] proposed an intelligent IoT-based system to diagnose and forecast Wheat diseases, pests and weeds. Another IoT-based system has been developed to manage the use of insecticides and fungicides based on predication models of crop disease and pest data [39]. Chieochan *et al.* [40] proposed an IoT-based irrigation system in the Lingzhi mushroom farm. They used humidity sensors which recorded high values between 90 to 95% humidity data in the farm. A Closed-Circuit Television (CCTV) system was established to monitor and control the sprinkler and fog pumps automatically.

**1.2 Contributions of this study**

In this article, we propose a wireless sensor network to sense and send a range of environmental parameters, namely temperature, humidity and soil moisture in different regions of a farm using Internet of Things (IoT) platform and a drone to collect all environmental recorded values and send them to the cloud-based server. Two main contributions of this study can be summarized as follows. First, we introduce a new water irrigation system by integrated the use of the UAV with IoT-based embedded system for smart farming. Second, the proposed irrigation system bas been practically implemented and tested in the laboratory successfully.

The remainder of this paper is divided into the following sections: Section 2 describes the proposed agricultural water irrigation system, including hardware and software components of the drone and IoT-based embedded system. Section 3 presents the practical results and evaluation of the developed smart irrigation system. Finally, conclusion and future remarks of this research work are given in Section 4.

## 2. Materials and Methods
### 2.1 Materials

Technical description of each component of our proposed agricultural UAV is illustrated in Table 1. It includes one Arduino microcontroller board, F330 4-Axis RC Quadcopter frame kit, four brushless motors with electronic speed control (ESC), radio frequency (RF) transmitter and receiver, and rechargeable LIPO battery. Moreover, schematic diagram of connected wiring components of the UAV is depicted in Fig. 3. A battery eliminator circuit (BEC) was added to supply electrical voltage to other circuits without using additional batteries.

**Table 1** Technical description of proposed UAV components in this study.

| Component | Description |
|---|---|
| Arduino UNO board | The Arduino UNO broad includes Atmel microcontroller with analog and digital input-output ports as a central unit for connecting and controlling drone hardware components altogether. |
| Drone frame | the drone frame in this study is F330 4-Axis RC Quadcopter Frame Kit. It has built-in circuit to support ESC connections. |
| Electronic Speed Control (ESC) | An electronic speed control or ESC is an electronic circuit that controls and regulates the speed of an electric brushless motor. |
| Brushless motor | Have been used in this project a servo motor is a rotary actuator or linear actuator that allows for precise control of angular or linear position, velocity and acceleration. |
| Rechargeable battery | LIPO battery is a rechargeable battery of lithium-ion technology using a polymer electrolyte instead of a liquid electrolyte. These batteries provide higher specific energy than other lithium battery types and are used in our designed drone system because of its light weight. |
| RF transmitter and receiver | The flight controller works by sending a radio signal from the remote control to the drone. That allows the drone to cover possible regions of the farm, collecting all required environmental data for smart irrigation. |

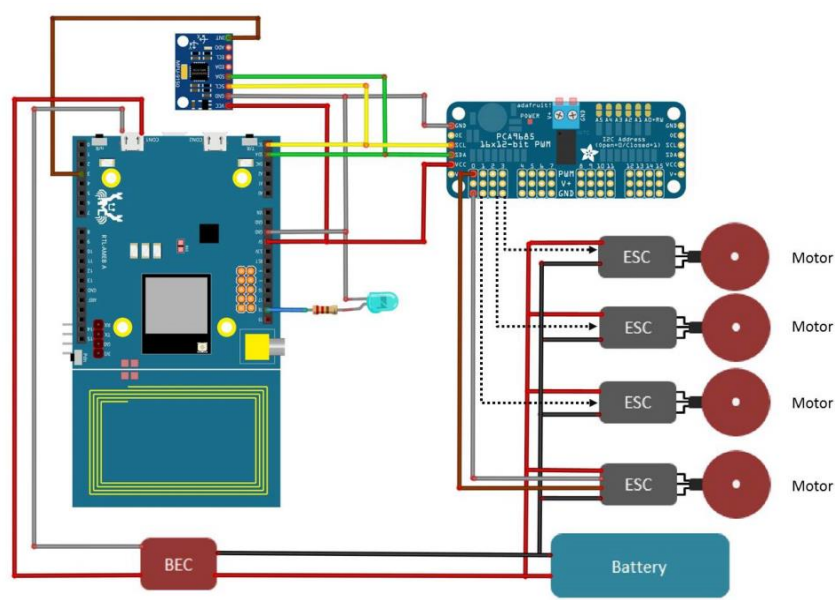

**Fig. 3** Schematic diagram of wiring circuit components of the proposed agricultural drone. All nomenclatures have been defined in the context.

For implementing our practical IoT-based irrigation embedded system, the following hardware and software components have been used for measuring the required environmental data and sending them via the drone to the cloud hosting server to handle the functional status of water pumps automatically.

- Arduino UNO microcontroller board
- Wi-Fi Module (ESP8266)
- DHT22 as temperature and humidity sensors
- Soil Moisture Sensor (FC28)
- Water pump 5.0 Vdc

The microcontroller board connects all environmental sensors of temperature and humidity sensors (DHT22) and soil moisture sensor (FC28) through analog input ports in the range of 5.0 Vdc. The measurement of soil moisture is calculated [41], as given by

$$Soil\ moisture(\%) = \frac{weight\ of\ water\ contained\ in\ the\ soil}{weight\ of\ dry\ soil\ samples} \tag{1}$$

In each region of the farm, the Wi-Fi module (ESP8266) is used for wireless communications (transmitting and receiving the environmental data) between the Arduino UNO board and the cloud server to automatically control the ON/OFF operations of the water pumps. The Arduino IDE is the development interface for programming the Arduino bord with peripherals, i.e. Wi-Fi module, electronic sensors and water pumps using ArduinoC programming language [42]. Arduino RemoteXY application provides a simple Android User Interface (UI) programming with free cloud services to remotely monitor the connected devices and the acquired data for the users [6].

**2.1. Methods**

Fig. 4 shows the schematic diagram of overall designed system components of our proposed IoT-based agricultural irrigation system. The drone is flying and collecting the data from two regions in the farm. Three readings of the environmental data, namely temperature , humidity and soil moisture are acquired by electronic sensors for each farm region. The Wi-Fi module sends these reading to the drone, which transmits them to the cloud. Then, the specialists or farmers can be notified by the status of the farm regions with the automatic estimation of water irrigation quantity. Finally, the developed system results will automatically operate the water pumps if needed.

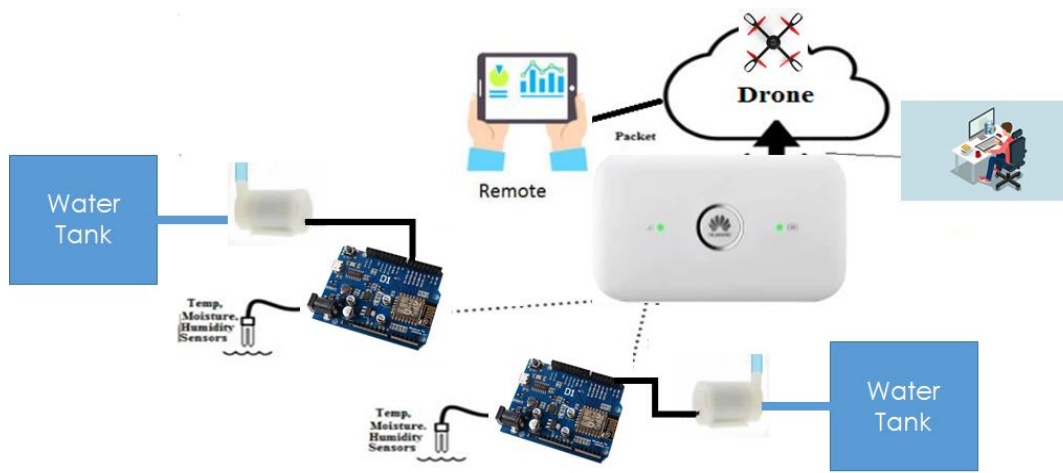

**Fig. 4** Schematic diagram of the proposed smart irrigation system using IoT and a drone.

The activity diagram of our proposed smart IoT-based irrigation system is shown in Fig. 5. The drone is initially configured with an Internet Protocol (IP) address for wireless connection with each Arduino microcontroller board in all regions of the farm. Once the drone covers a farm region, it collects the corresponding data of temperature, humidity and soil moisture, and send them to the cloud for getting the decision of operating the water irrigation pumps.

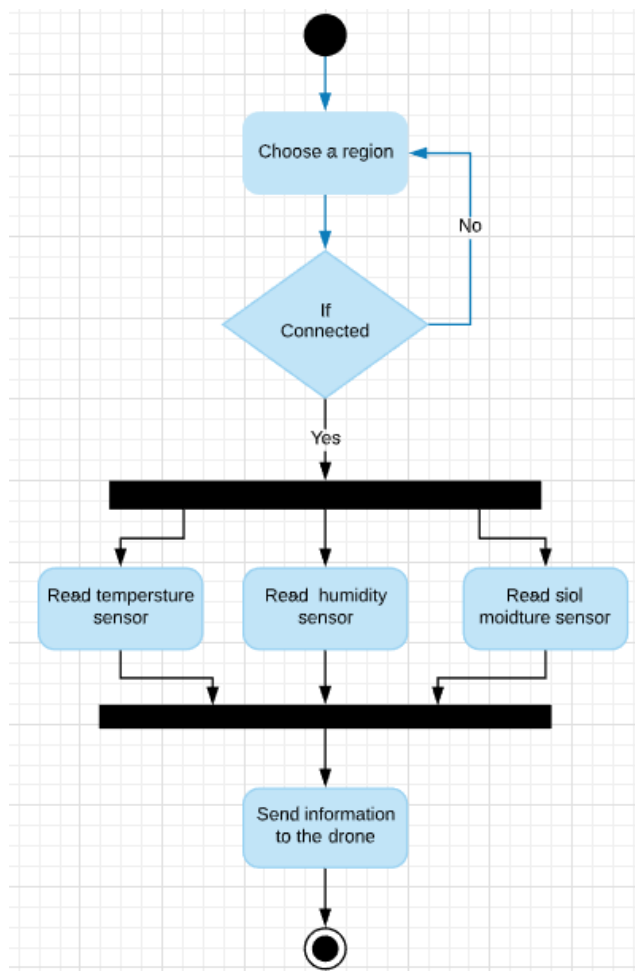

**Fig. 5** The activity diagram of monitoring the environmental data of temperature, humidity, and soil moisture using a drone and IoT-based system.

### 3. Results and Discussion

Fig. 6 shows the developed Android-based mobile user interfaces of our smart irrigation system using Arduino RemoteXY. In this study, two regions of the farm are defined as Region 1 and Region 2. The farmer can simply select any agricultural region and display the corresponding environmental data. The measured values of temperature (º C), humidity (%) and soil moisture (%) are monitored per region in the IoT-cloud system. The processing latency of estimating the quantity of water irrigation is 0.5 second because of the wireless delay transmission of the above measured data from sensors. Fig. 7 depicts the practical prototype of our developed smart irrigation system. Moreover, the hardware components of our proposed UAV for collecting environmental data in the farm are given as screenshots in Fig. 8. Furthermore, the environmental sensor readings have been successfully tested and evaluated and compared with the real weather news for temperature and humidity data in Riyadh, Saudi Arabia, as shown in Fig. 9. The successful water irrigation control system using artificial neural networks has been previously assessed in our study, as presented in Ref. [6]

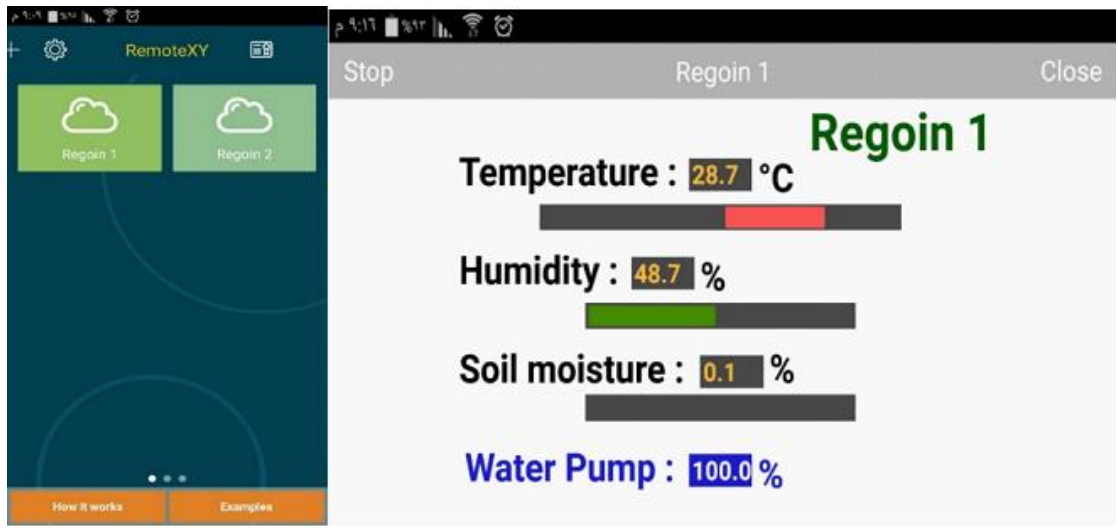

**Fig. 6** Mobile user interfaces (UI) of the developed IoT-based water irrigation system based on the measurements of temperature, humidity and soil moisture to operate the corresponding water pump.

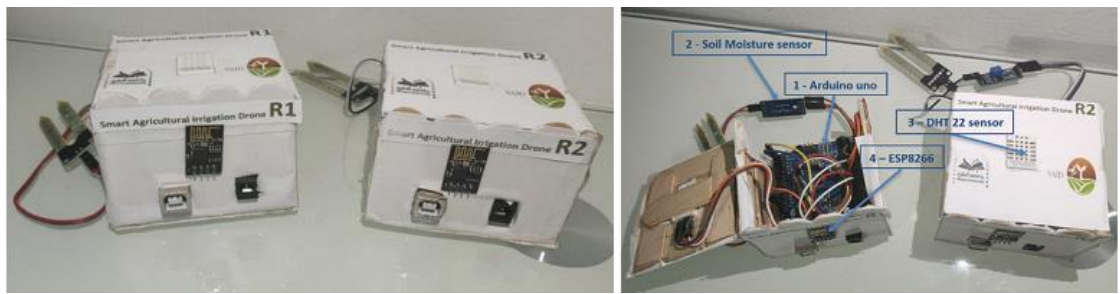

**Fig. 7** Practical implementation of the developed IoT-based irrigation system including Arduino board, Wi-Fi module, electronic sensors of temperature, humidity and soil moisture.

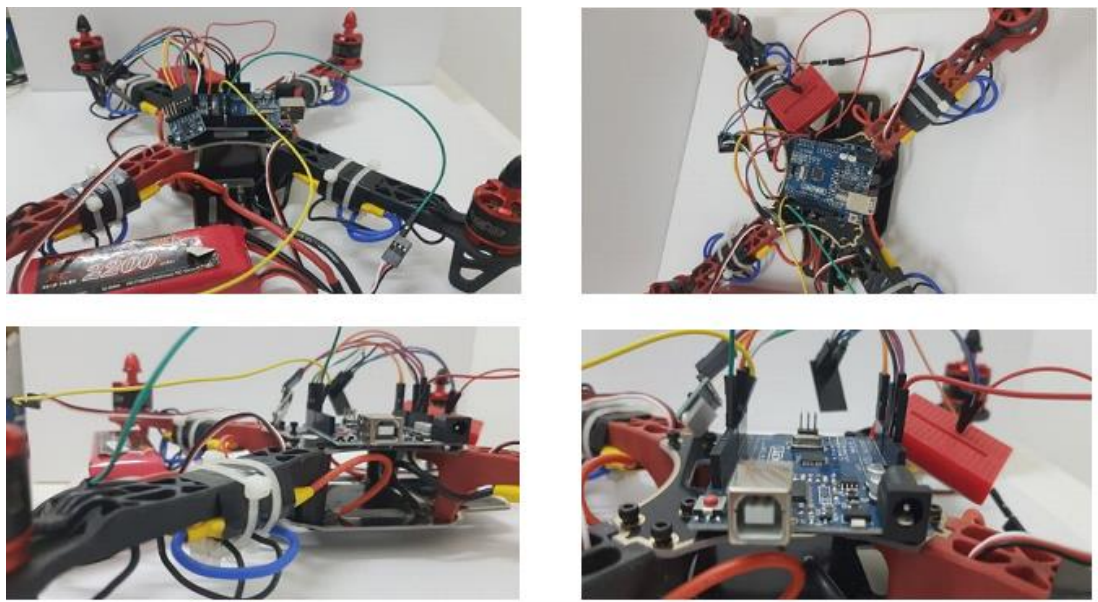

**Fig. 8** Screenshots of the practical design of our proposed agricultural UAV.

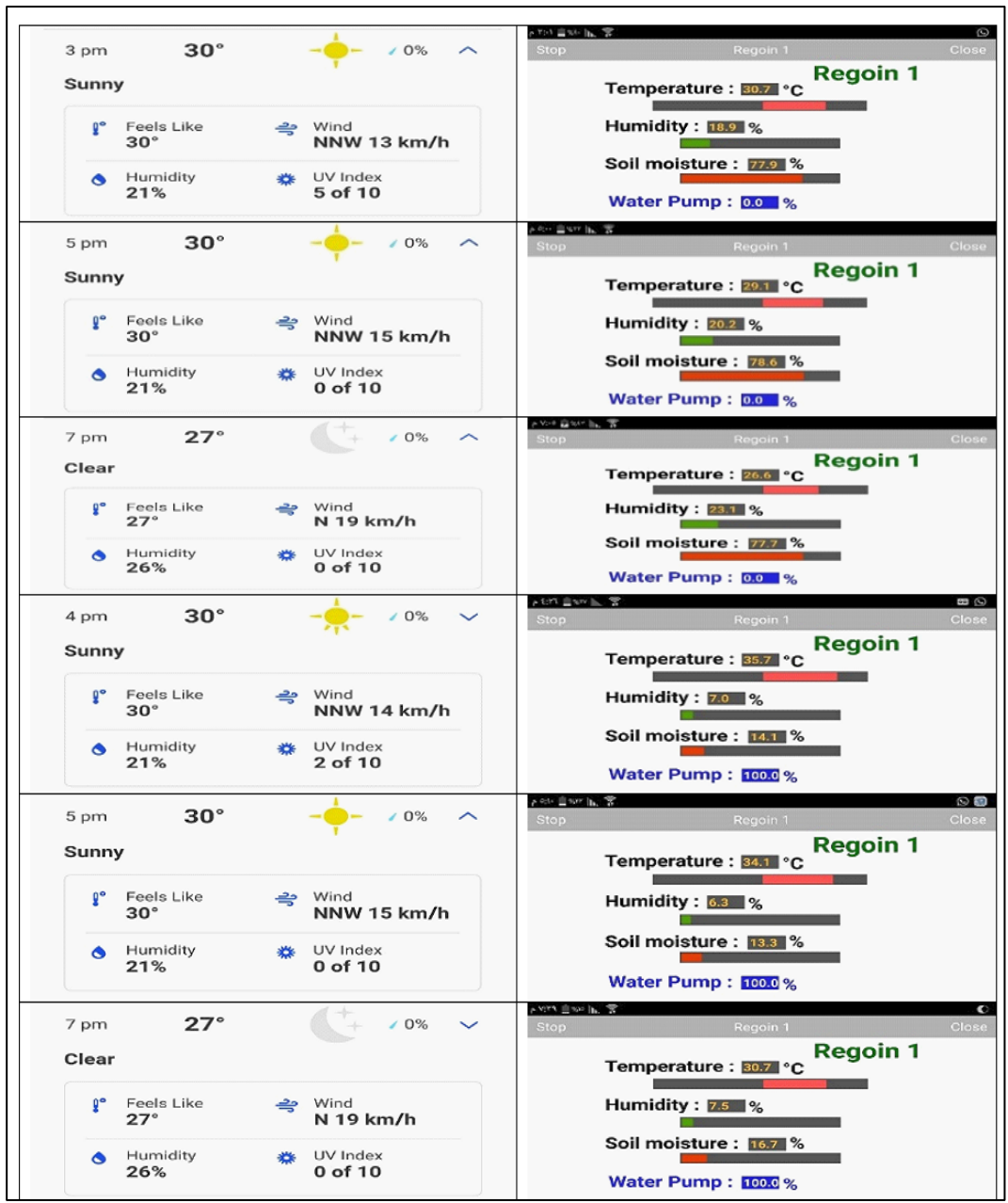

**Fig. 9** Comparative readings between the weather news and the developed system sensor readings
[Location: Riyadh; Date: 15 July 2020; Time: 3:00 -7:00 PM]

Although our developed IoT-based embedded system showed good results to monitor the environmental data of temperature, humidity and soil moisture to control the water irrigation pumps in two regions, as shown in Fig. 6 and 9. This study is still limited to experiments in the laboratory. The proposed agricultural drone is still under development to accomplish the tasks of collecting data in open air of the farm with permissions of official authorities and stakeholders in Saudi Arabia. Nevertheless, the designed and development of our smart irrigation system is still valid and scalable for saving water resources in the farm using IoT and UAV.

## 4. Conclusions and Outlook

In this study, a new smart IOT-based irrigation system has been successfully implemented to monitor the environmental data in the agricultural field. In addition, a drone has been designed and partially implemented to support smart water irrigation system, as shown in Fig. 8. Therefore, the main prospect of this research work is the real-time deployment of our developed irrigation system in the farms around the province area of Shaqra, Saudi Arabia. Also, this smart irrigation system can be extended to accomplish the following advanced tasks:1) Exploring water irrigation needs to a specific crop and inform the farmer by SMS; and 2) Analysis of crop growth stages using an UAV-based camera. Furthermore, Automatic flight control using the hand movement will be considered in the future design of our proposed agricultural UAV.

**Acknowledgment**

This project is supported by the College of Computing and Information Technology, Shaqra University.

[36] P. Zhang, Q. Zhang, F. Liu, J. Li, N. Cao, and C. Song, "The Construction of the Integration of Water and Fertilizer Smart Water Saving Irrigation System Based on Big Data," in *2017 IEEE International Conference on Computational Science and Engineering (CSE) and IEEE International Conference on Embedded and Ubiquitous Computing (EUC)*, 2017, vol. 2, pp. 392-397.

[37] T. Oksanen, R. Linkolehto, and I. Seilonen, "Adapting an industrial automation protocol to remote monitoring of mobile agricultural machinery: a combine harvester with IoT," *IFAC-PapersOnLine,* vol. 49, no. 16, pp. 127-131, 2016/01/01/ 2016.

[38] S. Zhang, X. Chen, and S. Wang, "Research on the monitoring system of wheat diseases, pests and weeds based on IOT," in *2014 9th International Conference on Computer Science & Education*, 2014, pp. 981-985.

[39] H. Lee, A. Moon, K. Moon, and Y. Lee, "Disease and pest prediction IoT system in orchard: A preliminary study," in *2017 Ninth International Conference on Ubiquitous and Future Networks (ICUFN)*, 2017, pp. 525-527.

[40] O. Chieochan, A. Saokaew, and E. Boonchieng, "IOT for smart farm: A case study of the Lingzhi mushroom farm at Maejo University," in *2017 14th International Joint Conference on Computer Science and Software Engineering (JCSSE)*, 2017, pp. 1-6.

[41] Radi, Murtiningrum, Ngadisih, F. S. Muzdrikah, M. S. Nuha, and F. A. Rizqi, "Calibration of Capacitive Soil Moisture Sensor (SKU:SEN0193)," in *2018 4th International Conference on Science and Technology (ICST)*, 2018, pp. 1-6.

[42] M. Böhmer, *Beginning Android ADK with Arduino*. Apress, 2012.